\documentclass[num-refs]{wiley-article}
\usepackage{siunitx}

\papertype{Original Article}

\title{Evaluation of distortion correction methods in diffusion MRI of the spinal cord}

\author[1]{Haykel Snoussi PhD}
\author[2]{Julien Cohen-Adad PhD}
\author[1]{Olivier Commowick PhD}
\author[1]{Benoît Combès PhD}
\author[1,3]{Élise Bannier PhD}
\author[4]{Anne Kerbrat MD, PhD}
\author[1]{Christian Barillot PhD}
\author[1]{Emmanuel Caruyer PhD}
\author[ ]{the EMISEP study group}

\affil[1]{Univ Rennes, CNRS, Inria, Inserm, 
          IRISA UMR 6074, Empenn ERL U 1228, Rennes, F-35000, France}
\affil[2]{Department of Electrical Engineering, 
          École Polytechnique de Montréal, Montréal, Québec, Canada}
\affil[3]{CHU Rennes, Radiology department, Rennes, F-35033, France}
\affil[4]{CHU Rennes, Neurology department, Rennes, F-35033, France}

\corraddress{Emmanuel Caruyer PhD, Univ Rennes, CNRS, Inria, Inserm, 
             IRISA UMR 6074, Empenn ERL U 1228, Rennes, F-35000}
\corremail{Emmanuel.Caruyer@irisa.fr}

\fundinginfo{Haykel Snoussi was partly funded by the EMISEP PHRC (ClinicalTrials.gov identifier NCT02117375), the Britanny region and a MITACS-Inria Globalink travel grant. MRI data acquisition was supported by the Neurinfo MRI research facility, University of Rennes 1. Neurinfo is granted by the the European Union (FEDER), the French State, the Brittany Council, Rennes Metropole, Inria, Inserm and the University Hospital of Rennes. Images of the healthy volunteers were acquired in the framework of OSS-IRM (ClinicalTrials.org identifier NCT03440983)}

\runningauthor{H Snoussi {\it et al.}}

\begin{document}
\maketitle

\begin{abstract}

\textbf{Background}: Magnetic field inhomogeneities generate important geometric 
distortions in reconstructed echo-planar images. Various procedures were 
proposed for correcting these distortions on brain images; yet, few 
neuroimaging studies tailored and incorporated the use of these techniques in 
spinal cord diffusion MRI.\\
\textbf{Purpose}: We present a comparative evaluation of distortion correction methods
that use the reversed gradient polarity technique on spinal cord. We propose
novel geometric metrics to measure the alignment of the reconstructed diffusion
model with the apparent centerline of the spinal cord.\\
\textbf{Subjects}: 95 subjects, among which 29 healthy controls and 66 multiple 
sclerosis patients.\\
\textbf{Sequence}: 3T, diffusion-weighted echo-planar imaging of the spinal cord, 
b-value 900s/mm$^2$, 30 directions.\\
\textbf{Assessment}: Geometric distortions were corrected using 4 state-of-the-art
methods. We measured the alignment of the principal direction of diffusion with
the apparent centerline of the spine after correction and the correlation with
the reference anatomical image. Results are computed per vertebral level, to
evaluate the impact on different portions of the spine. Besides, a
global, subjective evaluation of the quality of the correction of healthy
subjects images was performed by three expert raters.\\
\textbf{Statistical tests}: A paired Tukey test associated to no difference between
metrics from the non-corrected data and metrics from each of the 4
correction methods were computed. A logistic model including a rater effect was fitted to the subjective evaluation data.\\
\textbf{Results}: As a result of distortion correction, the diffusion directions are
better aligned locally with the centerline ($p < 0.05$), in particular at
both ends of the acquisition window. The cross-correlation with anatomical
image is also improved by HySCO ($p=2.10^{-4}$) and block-matching ($p=0.029$). The subjective evaluation for HySCO is significantly better (p < 0.05) than for Block-Matching; TOPUP performs significantly worse than the three other methods.\\
\textbf{Data Conclusions}: Correction based on Hyperelastic
Susceptibility Artefact Correction (HySCO) provide best results among the
selected methods.

\end{abstract}

\section{Introduction}
Echo-Planar-Imaging (EPI) is a fast and efficient magnetic resonance imaging
(MRI) acquisition technique \cite{stehling1991echo} useful for
diffusion-weighted MRI. It is however sensitive to the $B_{0}$ field
inhomogeneities, caused by the local susceptibility variations in the human
body. In this respect, \textit{in vivo} spinal cord imaging is especially
challenging due to the presence of various surrounding tissues and
environments: bone, cerebrospinal fluid (CSF), gray and white matter, muscle,
fat and air. Besides, the elongated shape of the spinal cord anatomy requires a
large field of view in the superior-inferior direction and its small
cross-sectional size requires high spatial resolution. These requirements
extend the readout durations in EPI which in turn distorts and blurs the image
along the phase-encoding direction (PED). Also, a large echo-spacing, the
interval between two consecutive line readouts in $k$-space, creates
distortions due to $B_{0}$ field inhomogeneity.

Geometric distortions affect the reconstructed image so that a voxel normally
at position $\mathbf{r}$ appears at a position $\mathbf{r} +
\mathbf{d_{\mathrm{PED}}}(\mathbf{r})$. This displacement,
$\mathbf{d_{\mathrm{PED}}}(\mathbf{r})$, occurs mainly in the PED and is
negligible in other directions
\cite{jezzard1995correction,macdonald2016efficient}.  Depending on parameters
of the EPI sequence such as the echo spacing, the field of view in the PED and
the $B_0$ field inhomogeneity), the magnitude of the displacements can reach up
to 60 mm \cite{saritas2014susceptibility} in typical spinal cord images (6ppm
off-resonance at 3T, field of view of \SI{18}{\centi\meter} in the PED and
echo-spacing of \SI{0.5}{\milli\second}) need to be addressed carefully before
analysis. In addition to the spatial distortion presented above, there also is
an intensity distortion in EPI \cite{jezzard1995correction}.

Various acquisition-based optimizations have been proposed to reduce
susceptibility artifacts for spinal cord imaging
\cite{saritas2014susceptibility}. In addition, there is a number of distortion
correction techniques, including co-registration, field map correction, point
spread function (PSF) and reversed gradient polarity method (RGP). Most of
these techniques require an additional specific acquisition and are based on
the distortion model derived in \cite{chang1992technique}.

First, among these techniques, \textit{co-registration} uses a non distorted anatomical image, typically a T$_2$-weighted acquisition as a reference.  Applying non-linear registration, a displacement field is computed to warp the non diffusion-weighted image to the T$_2$-weighted image. The same transform is then applied to the diffusion-weighted images prior to diffusion model fitting.  In general, imaging protocols do include an anatomical scan so this technique does not increase the overall acquisition time. However, this technique is sensitive to the quality of the registration, a task which can be challenging due to differences in contrast and field of view between the diffusion and the anatomical image.

The second technique, \textit{field map correction}, relies on a 2 minutes acquisition of a double-echo gradient-echo sequence, from which a $B_0$ field map is reconstructed \cite{jezzard1995correction, reber1998correction}. This map is in turn used to generate a warping field which will be applied to the EPI series. This technique is convenient, but is sensitive to subject motion during acquisition as well as large local variations in the magnetic field \cite{andersson2003correct,holland2010efficient}. Post-processing using this kind of acquisition is implemented in FSL\footnote{\url{https://fsl.fmrib.ox.ac.uk/fsl/fslwiki/FUGUE/Guide}} and \textit{Statistical Parametric Mapping} (SPM) \footnote{\url{https://www.fil.ion.ucl.ac.uk/spm/}} toolbox.

Another technique \cite{robson1997measurement,zeng2002image} also relies on the acquisition of an extra dedicated sequence from which the \emph{point spread function} of the system is estimated. The technique can be accelerated using parallel imaging \cite{zaitsev2004point,speck2008high}. The point spread function mapping captures both intensity and geometric distortions induced by field inhomogeneity. Applying this technique to diffusion MRI of the spinal cord  \cite{lundell2013fast} was proven effective in recovering large distortion, thus enabling fiber tractography in the cervical region.
    
The last technique, commonly referred to as \textit{reversed gradient polarity} (RGP), uses two EPI acquired with the same field of view and matrix size, but in opposite phase encoding direction \cite{chang1992technique}. The apparent displacements in these reconstructed images are of the same magnitude but have opposite directions. One can exploit this to compute the corresponding displacement map using both images as input: the corrected image is therefore midway between the two distorted acquisitions. Some studies consider this technique as an alternative to \textit{field map correction} as the field map here is implicitly estimated using an additional EPI. This RGP technique is distinctive by requiring only a quick additional EPI acquisition to correct distortion, if we consider that acquiring only the non diffusion-weighted image in reverse PED is sufficient. Furthermore, Cohen-Adad et al. shown its potential for the correct estimation of both geometric and intensity distortion as compared to \textit{PSF}, \textit{field map} and \textit{co-registration} \cite{cohen2009distortion}. 

In this work, we focus on the evaluation of post-processing distortion correction techniques using the RGP technique. We propose a comparative evaluation of post-processing methods using the pair of acquisitions for the correction of geometric distortions.  We first give an overview of the state-of-the-art methods we will evaluate. Then, we introduce a new geometric measure of alignment between the centerline of the spinal cord and the reconstructed principal direction of diffusion. Finally, we compare these methods on a database of 95 subjects (29 healthy controls and 66 multiple sclerosis patients) using the newly introduced measure of alignment, together with a comparison to the anatomical T2-weighted image using cross-correlation and mutual information. In addition to these objective evaluations, a subjective comparison of the images corrected with each of the 4 methods was performed by 3 independent raters. To the best of our knowledge, this is the first time such a comparison of these methods, otherwise designed for brain imaging, is made for spinal cord imaging.

\section{Materials and methods}
\subsection{Distortion correction from RGP technique}
A range of post-processing methods have been proposed for the estimation and the correction of geometric distortions in EPI from a pair of acquisitions with RGP. In this work we compare 4 such distortion correction methods: block-matching (BM) \cite{hedouin2017block}, hyper-elastic susceptibility artefact correction (HySCO) \cite{ruthotto2012diffeomorphic}, TOPUP \cite{andersson2003correct} and Voss \cite{voss2006fiber}. In this section, we first present the physical model common to all methods, then we give an overview of the image processing methodology for each.

\subsubsection*{Distortion model}
A pair of spin-echo EPI are acquired in reversed phase encoding directions, in our case head-feet (HF) and feet-head (FH), which correspond to the $y$-axis. Let $I_F$ denote the forward gradient EPI image, acquired traversing $k$-space in the positive $y$-direction (HF), and $I_B$ (for backward) the reverse gradient image acquired traversing in the negative $y$-direction FH. The corrected image $I_C$ is therefore midway between $I_F$ and $I_B$. The problem is to find the deformation field from $I_F$ (or equivalently from $I_B$) to $I_C$ \cite{hedouin2017block}.

\subsubsection*{Method 1: Block-Matching}
The block-matching (BM) strategy looks for a transformation by finding a local correspondence between small blocks (typically $3\times 3 \times 3$ voxels) in the original and the target image \cite{ourselin2000block, commowick2012automated}. This method, applied to field inhomogeneity-induced distortion correction in EPI \cite{hedouin2017block}, follows the registration method introduced in \cite{avants2008symmetric}. BM \cite{hedouin2017block} adapts the approach to block matching algorithm presented by \cite{ourselin2000block,commowick2012automated} by constraining the transform to be aligned with the PED, using a cross-correlation metric as a measure of similarity. BM method is implemented and available in Anima\footnote{\url{https://github.com/Inria-Visages/Anima-Public/}}, an open-source software for medical image processing.

\subsubsection*{Method 2 : HySCO}
The Hyper-elastic Susceptibility artefact COrrection method (HySCO) \cite{ruthotto2012diffeomorphic,macdonald2016efficient} is based on the physical distortion model derived in \cite{chang1992technique}. HySCO estimates the displacement $b: \Omega \to \mathbb{R}$ by minimizing the following distance functional:
\begin{equation}
  \mathcal{D}[b] = \frac{1}{2} \int_\Omega \left( 
    I_F(\mathbf x + b(\mathbf x)\mathbf v) (1 + \mathbf v^\top \nabla b(\mathbf x)) -  
    I_B(\mathbf x - b(\mathbf x)\mathbf v) (1 - \mathbf v^\top \nabla b(\mathbf x))
  \right)^{2} \mathrm{d}\mathbf x
  \label{eq:eq_hysco1}
\end{equation}
\noindent
where $\mathbf v$ denotes the PED, $\mathbf v = (0,1,0)^\top$, $\Omega$ the rectangular region of interest, $b(\mathbf x)$ is the magnitude of the field inhomogeneity at $\mathbf x \in \Omega$. Since minimizing $\mathcal{D}$ is an ill-posed problem, HySCO introduces the following regularization:
\begin{equation}
  \mathcal{S}(b) = \frac{1}{2} \int_\Omega 
                   \left\Vert \nabla b(\mathbf x) \right\Vert^2 \mathrm{d}x.
  \label{eq:eq_hysco3}
\end{equation}

HySCO is implemented as a part of ACID \footnote{\url{http://www.diffusiontools.com/}} toolbox and integrated as a batch tool in the \textit{Statistical Parametric Mapping} (SPM).

\subsubsection*{Method 3: TOPUP}
Similarly, TOPUP \cite{andersson2003correct} takes as input two images with opposite PEDs and jointly estimates a displacement field map and an undistorted image using least squares, to be used to reconstruct the undistorted image. TOPUP is implemented and distributed as part of the FSL package \footnote{\url{https://fsl.fmrib.ox.ac.uk/fsl/fslwiki/topup}}. Notice that when applying TOPUP algorithm on data, extreme slices of image volume are collapsed. We applied TOPUP with its default parameters except for resampling, we used the Jacobian method. 

\subsubsection*{Method 4: Voss}
Voss et al. \cite{voss2006fiber} proposed a distortion correction method to reduce geometric and intensity distortions. For each image coordinates $x$ and $z$, this method seeks to align the corresponding line in the forward and backard images at coordinates $x, \cdot, z$. Note that this method realigns each line independently, so in order to increase the coherence between adjacent lines, a 3D Gaussian smoothing is applied to the deformation. This method is implemented and available in Anima\footnote{\url{https://github.com/Inria-Visages/Anima-Public/}}. We used the method with default parameters in our comparison.

\subsection{Geometric measure of alignment} \label{subsec:geo_measure}
In order to evaluate the impact of geometric distortion correction, we introduce in this section a measure of alignment between the geometry of the spinal cord, represented by its centerline, and the principal direction of diffusion. In previous studies \cite{lundell2013fast,voss2006fiber,cohen2009distortion}, one of the criteria used for the evaluation or the validation of distortion-correction was to use fiber tractography in the spinal cord. We hypothesize that measuring the alignment of the diffusion direction and the spinal cord geometry is an alternative, more direct way to evaluate the consistency between the diffusion anisotropy and the underlying image geometry.

\begin{figure}[ht]
\begin{center}
    \includegraphics[width=.7\textwidth]{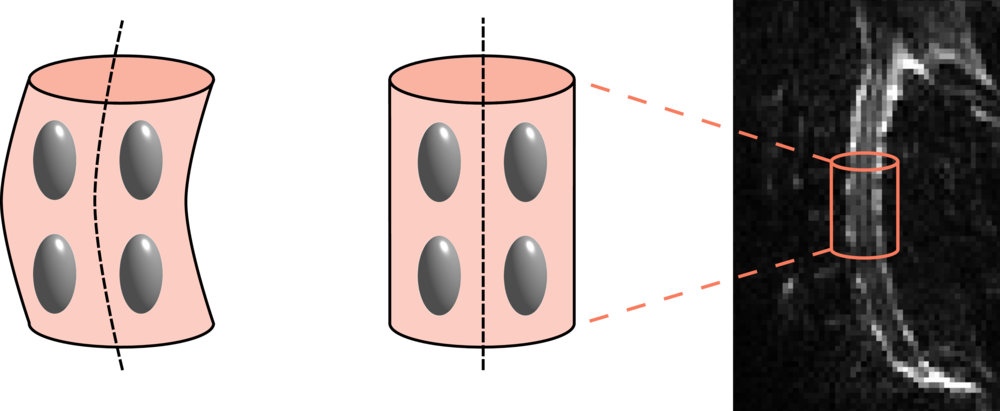}
  \caption{Illustration of the diffusion tensor field within the spinal cord: distorted image (left), undistorted (middle). The principal direction of diffusion is not affected by distortion, as a result it appears misaligned with the centerline of the spine on the distorted image (left).\label{fig:spine_rep}}
\end{center}
\end{figure}

The diffusion tensor model relates the local displacement of water molecules with the surrounding tissue microstructure. In the spinal cord white matter, it is mainly aligned with longitudinal fibers \cite{cohen2008detection}, which themselves follow a path parallel to the centerline of the spinal cord. We can therefore expect that the principal eigenvector of the diffusion tensor is locally aligned tangentially with the centerline of the spinal cord. However, when the image is distorted, the apparent shape of the spinal cord is affected, while the direction of the tensor is not. This results in a local misalignment of the diffusion tensor with the spinal cord (see Figure.\ref{fig:spine_rep}). In what follows, we describe a method to measure how the diffusion tensors and the centerline of spinal cord align with each other. 

\subsubsection*{Centerline extraction and modeling}
We first segment the diffusion image to obtain a binary mask of the spinal cord using the spinal cord toolbox (SCT) \cite{de2014robust,de2015automatic}.  This segmentation is computed from the mean diffusion-weighted images. Using this binary mask of the spinal cord, we compute the centerline in two steps: first, for every axial slice, we compute the barycenter of the mask within this slice; then we fit a degree-3 smoothing spline to the set of barycenters, using \texttt{UnivariateSpline} provided in SciPy\footnote{\url{https://www.scipy.org/}} with default smoothing parameter. This provides us with a continuous, differentiable representation of the centerline as a parametric equation of variable $t$, from which we can compute the Frenet frame at every point of the centerline.
\subsubsection*{Local frame attached to the spine centerline}
Let $\mathbf x(t) \in \mathbb{R}^3$ be a curve parametrized by $t \in \mathbb{R}^+$. Also, let $\mathbf {v(t)} = \mathbf{x(t)}'$ denotes the velocity. We define the curvilinear abscissa $s$ as:
\begin{equation}
  s(t) = \int_0^t |\mathbf x'(\tau)| \mathrm{d}\tau,
\end{equation}

\begin{figure}[ht]
\begin{center}
  \includegraphics[width=\textwidth]{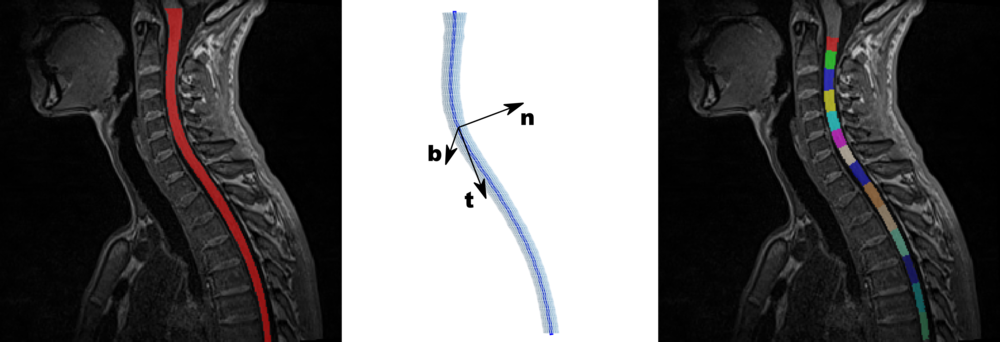}
  \caption{(left) T1 image with segmented spinal cord for reference; (middle)
  3d representation of the segmented spine, smoothed centerline (blue solid line) and the
  Frenet frame at a particular point of the centerline; (right) registration of the PAM50 template
  to identify intervertebral levels of the spine.\label{fig:frenet-fit}}
\end{center}
\end{figure}

The Frenet frame is defined as the triplet $(\mathbf t, \mathbf n, \mathbf b)$ where $\mathbf t$ is the tangent, $\mathbf n$ the normal and $\mathbf b$ the binormal (see Figure.\ref{fig:frenet-fit}) defined as 
\[
\mathbf{t} = \frac{\mathbf{v(t)}}{\|\mathbf{v(t)}\|};\quad
\mathbf{n} = \frac{\mathbf t'}{\|\mathbf t\|};\qquad
\mathbf b = \frac{\mathbf t \times \mathbf n}{\|\mathbf t \times \mathbf n\|}.
\]

\subsubsection*{Measuring the alignment of the diffusion model with the cord}
For every voxel at position $\mathbf r$ within the spinal cord, we first compute the closest point $\mathbf x(t_0)$ to $\mathbf r$ along the centerline where $t_0$ is by minimizing (with an exhaustive grid search):
\begin{equation}
  t_0 = \arg\min_t \{ ||\mathbf x(t) - \mathbf r||\}.
\end{equation}
Then we compute the coordinates of the principal eigenvector of the diffusion tensor, $\mathbf e_1(\mathbf r)$, in the Frenet frame computed at $\mathbf x(t_0)$. 

In order to summarize the distribution of the principal eigenvectors $\mathbf e_1(\mathbf r)$ within a region $\Omega$, we compute the covariance matrix of these directions \cite{mardia2009directional} as follows:
\begin{equation}
  \mathbf M = \frac{1}{\mathcal V(\Omega)} \int_\Omega \mathbf e_1(\mathbf r)\mathbf e_1(\mathbf r)^\top \mathrm d\mathrm r
\end{equation}
where $\mathcal V (\Omega)$ is the volume of $\Omega$. Intuitively, the matrix $\mathbf M$ will characterize the statistics of the angular deviation between the diffusion direction and the tangent to the centerline. From the eigendecomposition of the symmetric matrix, $\mathbf M$, we extract two statistics:\\
\textbf{Mean angle direction (MAD)}: defined as the angle in degrees between the
  principal eigenvector of $\mathbf M$ and [1,0,0], which corresponds to
  $\mathbf T$ in the Frenet frame\\
\textbf{Angular concentration of directions (ACD):} defined as the first
  eigenvalue of $\mathbf M$. Being the average of rank-1 non-negative symmetric
  matrices with eigenvalues (1,0,0), $\mathbf M$ has eigenvalues in the range of
  [0,1]. The more concentrated around the mean direction, the closer to 1 is the ACD.

These statistics were computed for $\Omega \in \{$Brain stem, C1, \ldots, C7, T1, T2$\}$ corresponding to every cervical and thoracic vertebral levels in the acquisition window.

\subsection{Imaging data}
Data used in this evaluation were acquired on 3T scanners (Magnetom Verio and Skyra, Siemens Healthineers, Erlangen) at four sites in France:  Marseille, Rennes, Strasbourg and Montpellier as part of the EMISEP project \footnote{\url{https://clinicaltrials.gov/ct2/show/NCT02117375}}.

\subsubsection*{Patients and healthy volunteers}
Twenty nine healthy volunteers (mean age = 32.83$\pm$7.13, 18F/11M) and 66 MS patients (mean age = 32.20$\pm$6.30, 42F/24M) were recruited in the study after approval from the institutional review board. MS patients are early relapsing-remitting MS patients (scan within the first year following diagnosis), with a median Expanded Disability Status Scale (EDSS) score of 1.0 (range [0, 2.5]). All participants provided informed written consent. Table.~\ref{table:nb_subjects_discor} illustrates more details about MRI scanners, centers and demographics.

\begin{table}[ht]
  \begin{center}
    \begin{tabular}{|l|c|c|c|c|c|}
      \hline
      Center & Marseille & Rennes & Strasbourg & Montpellier & \textbf{TOTAL}\\
      \hline
      3T scanner MRI & Verio & Verio & Verio & Skyra & - \\
      \hline
      Volunteers & 4 & 18 & 3 & 4 & \textbf{29}\\
      \hline
      Gender & 1F/3M & 10F/8M & 3F & 4F & \textbf{18F/11M}\\
      \hline
      \hline
      MS Patients & 7 & 44 & 8 & 7 & \textbf{66} \\
      \hline
      Gender & 4F/3M & 28F/16M & 5F/3M & 5F/2M & \textbf{42F/24M} \\
      \hline
      \hline
      \multicolumn{5}{c|}{}&\textbf{95 (60F/35M)}\\
      \cline{6-6}
    \end{tabular}
    \vspace{0.75em}
    \caption{Demographic information for all participating subjects, healthy controls and MS patients from several centers and total study cohort.
    \label{table:nb_subjects_discor}}
  \end{center}
\end{table}
\subsubsection*{MRI data acquisition}
EMISEP PHRC protocol includes Diffusion-weighted imaging scan: Thirty non-collinear diffusion-weighted images (DWI) were acquired at $b = 900$~s$\cdot$mm$^{-2}$, six non-DWI ($b = 0$) measurements and one non-DWI ($b = 0$s/mm$^2$) with an opposite phase encoding direction (PED) were also acquired. This was repeated three times successively in order to increase the signal-to-noise ratio (SNR). Scans were performed in sagittal orientation and head-feet (HF) PED to cover the whole cervical cord. The pulse sequence used for diffusion MRI was a single-shot Echo-planar imaging (ss-EPI) using parallel imaging (GRAPPA, acceleration factor 2). Sixteen slices were acquired with the following parameters without inter-slice gap: TR/TE = 3600/90~ms, with 2x2x2~mm$^3$ as the resolution, and image matrix 80x80. The total acquisition time for the DWI sequence was approximately 7 minutes.
The protocol also includes the following two high-resolution images for anatomical reference. Firstly, T$_1$-weighted scan: in sagittal orientation, magnetization-prepared rapid acquisition gradient echo (3D MPRAGE) sequence with an isotropic 1x1x1~mm$^3$ resolution, TR/TI/TE = 1800/900/2.79~ms,  flip  angle 9$^\circ$ and FoV = 250$\times$250 mm$^2$, 64 slabs. Secondly, T$_2$-weighted scan: in sagittal orientation, 2D Turbo Spin Echo, with anisotropic 0.7x0.7x2.75~mm$^3$ resolution, TR/TE = 3000/68.0~ms and FoV = 260$\times$260 mm$^2$, 15 slices with 2.5 mm slice thickness and 10\% gap between slices.

\subsection{MRI pre-processing}
\subsubsection*{Motion correction} 
The diffusion-weighted volumes were re-aligned to compensate for subject motion during the 7 minutes acquisition. Since motion is often observed within the axial plane \cite{mohammadi2013impact}, the correction were carried out allowing only rigid slice-wise transformations in the axial plane using the spinal cord toolbox (SCT) version 3.0b4 \cite{xu2013improved,de2017sct}. This way of realigning all volumes was shown to be robust and accurate \cite{mohammadi2013impact}.

\subsubsection*{Distortion correction} 
Then, diffusion-weighted data were corrected for susceptibility distortion using the four cited methods: Block-Matching (BM) available in Anima, HySCO as implemented in SPM-ACID toolbox, TOPUP implemented in FSL and Voss method implemented in Anima. We computed a displacement field using one non-DWI ($b = 0$) for each PED. Then, this displacement was applied to the whole set of diffusion-weighted volumes. We refer to the diffusion-weighted images after distortion correction as "corrected DWI".

\subsubsection*{Segmentation of the cord and identification of vertebral level}
Using SCT \cite{de2017sct}, whole spinal cord segmentation was carried out on T$_{1}$-weighted, on T$_{2}$-weighted, on the corrected by each method and uncorrected DWI (b = 900~s$\cdot$mm$^{-2}$) averaged across all gradient directions. Then, we identify manually two vertebral levels, C3 and T1 since this manual step is recommended in the SCT for an accurate registration to the PAM50 template.  

\subsubsection*{Quality Control}\label{sec:qc_discor}
To ensure that the results are immune to image artefacts, and to identify problems during the processing pipelines, we performed a careful quality control (QC) on the raw data and after each processing step. We eliminated 7 subjects which contained too many artifacts not apparently related to B$_{0}$-field inhomogeneity (motion and ghosting). 

\subsection{Computing geometric metrics by vertebral level}
Here, our aim is to compute statistics on geometric measures, MAD and ACD, for each vertebral level.  To do so, we register the PAM50 template \cite{de2018pam50} and the T$_{1}$-weighted anatomical data and the mean diffusion-weighted image. Firstly, T$_{1}$-weighted anatomical data were registered to the PAM50 spinal cord template \cite{de2018pam50}. This generates forward and inverse warping fields from one to the other.  Next, we register the PAM50-T1 template \cite{de2018pam50} to the mean DWI using the inverse warping field from the previous step as an initial warping field. Then, using the registered template, we compute the proposed geometric metrics as explained in \ref{subsec:geo_measure} within each vertebral level.

\begin{figure}[ht]
\begin{center}
  \includegraphics[width=1.01\textwidth]{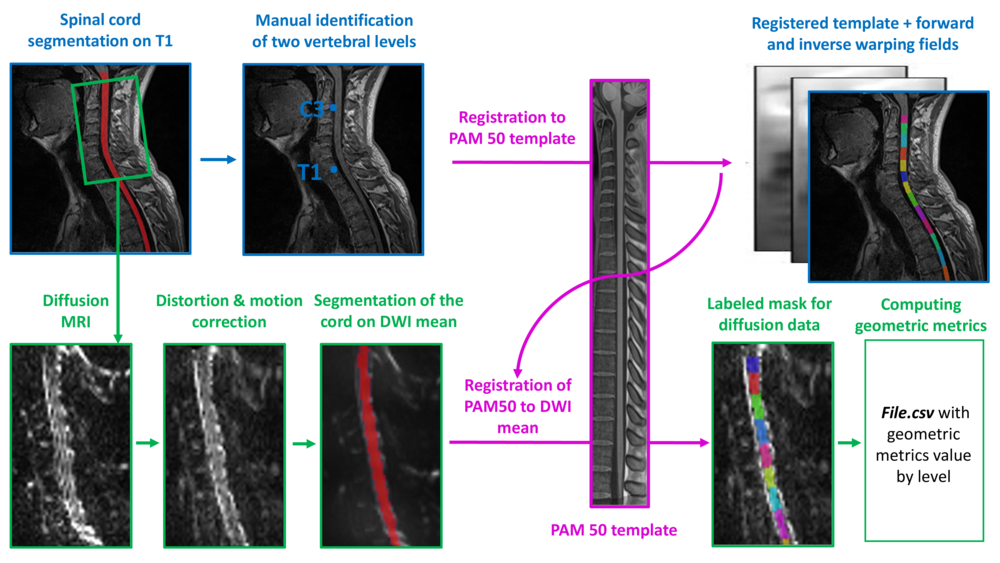}
  \caption{Illustration of the automated analysis pipeline. (1) Segmentation of
  the cord on T$_1$-weighted. (2) Manual identification of two vertebral levels. (3)
  Registration to the PAM50 template. (4) Motion and distortion correction of
  DWI. (5) Segmentation of the cord using DWI mean data. (6) Registration
of PAM50-T1 registered to DWI mean data using the inverse warping field from
previous registration as an initial warping field. (7) Computing MAD and ACD by vertebral level of the
cervical part.
  \label{fig:pipeline}}
\end{center}
\end{figure}

Here, we preferred to use T$_{1}$-weighted image rather than the T$_{2}$-weighted since the former has isotropic resolution, which makes the registration more effective. Alignment with the template provides robust definition of the inter-vertebral levels for the spinal cord. This enables computation of the average metrics in spinal cord using the atlas-based approach introduced in \cite{levy2015white}. As a result, we can quantify the geometric diffusion-based metrics averaged for each inter-vertebral level from C1 to T2. The processing pipeline is summarized in Figure.~\ref{fig:pipeline}.

\subsection{Computing cross-correlation and mutual information}
Complementary to the geometric measure of alignment, we also compute the similarity between the corrected $b=0$ image and the T$_2$-weighted image. Firstly, we rigidly registered T$_{2}$-weighted scan to the first $b=0$ volume of the uncorrected and corrected DWI. We then apply this rigid transform to the binary mask of the spinal cord obtained by segmenting the T$_{2}$-weighted image, which is now warped to the first $b=0$ volume of the uncorrected and DWI.  Finally, we compute the cross-correlation (CC) and mutual information (MI) between the T2-weighted image and the $b=0$ volume only within the spinal cord region. 
The cross-correlation, also referred to as Pearson correlation coefficient, is a measure of the linear correlation between two variables $x$ and $y$. This coefficient has a value in $[-1,1]$, where -1 and 1 are total negative and positive linear correlation respectively and 0 is no linear correlation. This coefficient is commonly represented by $r_{xy}$ and defined as:
\begin{equation}
  \mathit{r_{xy}} = 
  \frac{\sum_{i=1}^{n} (x_i - x_m)(y_i-y_m)}
       {\sqrt{\sum_{i=1}^{n}(x_i - x_m)^2 \sum_{i=1}^{n}(y_i-y_m)^2}}
  \label{cc_equation}
\end{equation}
\noindent
where $n$ is the size of the mask in voxels, $x_i$ and $y_i$ are the  intensities at voxel $i$ of the T$_{2}$-weighted and DWI, $x_m$ and $y_m$ are their sample mean over the same ROI.
The mutual information (MI) is a measure of the mutual dependence between two variables $x$ and $y$. It quantifies the amount of information obtained about $x$ through observing $y$. MI of $x$ and $y$ is given as:
\begin{equation}
  \mathit{MI(x,y)} = 
  \sum_{i=1}^{n} \sum_{i=1}^{n} p(x_i,y_i)~log 
  \left(
    \frac{p(x_i, y_i)}{p(x_i)p(y_i)}
  \right),
  \label{mi_equation}
\end{equation}
where $p(x,y)$ is the joint probability function of $x$ and $y$, and $p(x)$ and $p(y)$ are the marginal probability distribution functions of $x$ and $y$, respectively. 


\subsection{Subjective evaluation}
Three experts (E.B, S.L and A.K) compared the resulting mean diffusivity (MD)
images obtained after applying each of the four distortion correction methods
described above on the 29 healthy volunteers in the database. Using a web-based
interface implemented internally, the raters were provided with the images
corrected using each method side-by-side, along with the non-corrected image.
The raters were asked to rank the four corrected images, based on their
subjective appreciation of the resulting image. The methods were unknown to the
raters and the order in which images were presented were shuffled randomly for
every subject. The ranks were exported to a CSV file and analyzed with R.

\section{Results}
\label{sec:results}
\subsection{Geometric statistics}
\begin{figure}[ht]
\begin{center}
  \includegraphics[width=\textwidth]{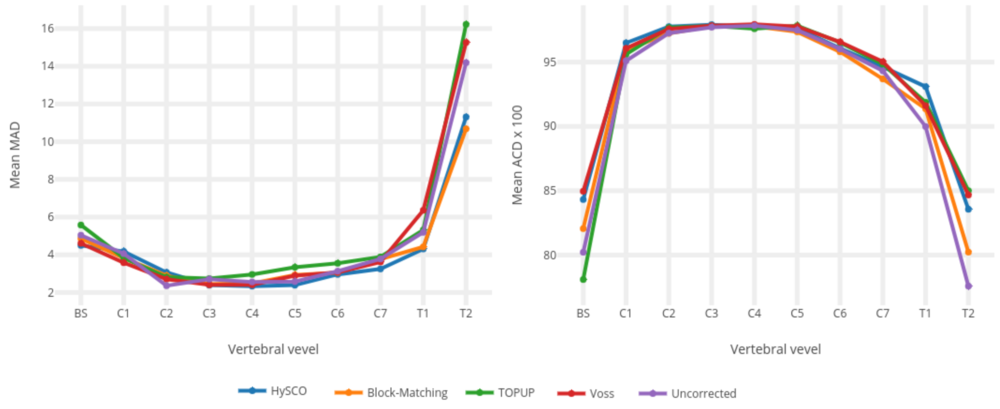}
  \caption{Mean MAD (in degrees) and mean ACD ($\times$ 100) reported by vertebral level. This way of representation shows that extreme vertebral level of the acquisition window are more affected by distortion.\label{fig:mean_mad_acd_plot}}
\end{center}
\end{figure}

For every distortion correction method, geometric statistics were computed for every cervical and thoracic levels in the acquisition window. Results of mean and standard deviation of each metric, MAD and ACD, for brain stem (BS) and each vertebral level are presented in tables ~\ref{table:mean_mad} and ~\ref{table:mean_acd}. For each level, the adjusted p-values associated to no difference between metrics from the non-corrected data and metrics from each of the 4 correction methods were computed using a paired Tukey test. Figure.~\ref{fig:mean_mad_acd_plot} shows the MAD and ACD by vertebral level; vertebral levels located at the end of the acquisition window are more affected by distortion. Note that MAD is a measure of the bias in the orientation, which means that a decrease in MAD should be interpreted as an improvement in the alignment. By contrast, ACD is a measure of angular dispersion, and an increase in ACD is considered as an improvement.

\subsection{Subjective comparison}
The relative performance of each correction method with respect to the others is summarized in \ref{table:subjective_table}. Moreover, for each pair of methods, a logistic model including a random rater-effect was fitted and the p-value for "log-odd=0" (i.e. in average method 1 performs better than method 2 in half of all virtually possible cases) was computed and reported. 

\subsection{Comparison with anatomical images}
Since the T$_{2}$-weighted image is not affected by distortion, it can be considered as a gold-standard for the evaluation of distortion correction. So, we also compare the corrected volume ($b=0$s/mm$^2$) of DWI by each method to the T$_{2}$-weighted image. We report results of cross-correlation and mutual information. Note that results presented here are slightly different to the study reported in \cite{snoussi2017comparison} since the treatment pipeline and datasets are different. A paired Tukey test was performed on the cross-correlation scores and reported in Table.~\ref{table:cross_corr} and on the mutual information scores and reported in Table.~\ref{table:mutual_info}. 

\subsection{Vertebral level volume}
Last, we computed the volume in mm$^3$ of each vertebral level for uncorrected and corrected data as shown in Table.~\ref{table:volume_level}. This table contains also results of paired Tukey test to compare the volume of each method to the non-corrected DWI. 

\section{Discussion}
\label{sec:Discussion}
In this work, we have proposed novel geometric diffusion-based metrics and framework for studying the impact of distortion correction in diffusion MRI of the spinal cord. We showed a difference in geometric alignment after correcting with one method or another. 

For HySCO method, as shown in Table.~\ref{table:mean_mad}, we can observe that MAD metric performs significantly better than uncorrected in T1 and T2 vertebral levels. But, this method has significant deterioration at C2 vertebral level, as BM, but mean MAD of HySCO is more close to the uncorrected. For ACD metric, we remark a significant improvement at edges of the spinal cord (brain stem, C1, C7 and T1 regions) as shown in Table.~\ref{table:mean_acd}. Last, we note as shown in Table.4 that HySCO performs significantly better in the subjective evaluation than BM and TOPUP. 

For Block-Matching method, as shown in Table.~\ref{table:mean_mad}, we observe that it performs significantly better in T1 vertebral levels. However, there is a significant deterioration in C2. For ACD metric, BM improves the concentration significantly in brain stem region, C1, C2, T1 and T2. We note also as shown in Table.~\ref{table:subjective_table} that BM performs significantly better in the subjective evaluation than TOPUP.

For Voss method, we observe a significant improvement in MAD metric for C1 vertebral level and significant deterioration in T1 vertebral level. But, there is a very remarkable amelioration for ACD in all vertebral levels, except C3 and C4 which are in the middle of the field-of-view of the acquisition. Last, we note as shown in Table.~Table.~\ref{table:subjective_table} that Voss performs significantly better in the subjective evaluation than TOPUP. However, TOPUP method is recognized as the major method for the postprocessing distortion correction of diffusion MRI data \cite{dauleac2021effect,hu2020distortion}.

For TOPUP method, we consider significant deterioration in MAD in C2 and C5 vertebral levels without any significant improvement. For ACD metric, there is significant improvement at C2, C5, C6 and T2 vertebral levels. Last, we note as shown in Table. \ref{table:subjective_table} that TOPUP performs significantly worse in the subjective evaluation than the other 3 methods.

Complementary to the geometric measure of alignment, we also compute the similarity between the corrected $b=0$ image and the T$_2$-weighted image. For cross-correlation, there is a significant amelioration only for Block-Matching and HySCO as demonstrated in Table.~\ref{table:cross_corr}. For mutual information, there is a significant amelioration for all methods as shown in Table.~\ref{table:mutual_info}. Last, we compared the apparent volume of each vertebral level before and after applying each distortion correction method (results reported in Table.~\ref{table:volume_level}. We note that all methods have a significant impact on the volume, which is related to the specific implementation of the distortion correction and interpolation methods. Despite having no ideal method to evaluate these changes, we decided to report them here, since they vary from one method to another.


\subsection{Conclusion}
We proposed a framework to evaluate distortion correction methods applied to diffusion MRI of the spinal cord. The results of objective and subjective evaluation on a database of 95 subjects show that among these methods, HySCO stands out in its ability to reduce the bias in diffusion direction (MAD) after distortion correction and in the subjective evaluation by experts. A possible next step in this research will be to incorporate this notion of alignment as a constraint to design a new correction algorithm tailored to spine diffusion MR images. 


\begin{table}
\begin{center}
\begin{tabular}{|c|c|c|c|c|c|c|c|c|c|c|}
  \hline
  \multicolumn{11}{c}{Mean Angle Direction (MAD)}\\
  \hline
  ~&\multicolumn{2}{c|}{\textbf{Block-Matching}}&\multicolumn{2}{c|}{\textbf{HySCO}}&\multicolumn{2}{c|}{\textbf{TOPUP}}&\multicolumn{2}{c|}{\textbf{Voss}}&\multicolumn{2}{c|}{\textbf{Uncorrected}}\\
  \hline
  \textbf{Level}&Mean&STD&Mean&STD&Mean&STD&Mean&STD&Mean&STD\\
\hline
BS &4.50 &3.59 &4.86 &3.48 &5.58 &4.61 &4.61 &3.76 &5.04 &3.94\\
\hline
C1 &4.18 &5.75 &3.83 &3.89 &3.84 &2.39 &\colorbox{green!25}{3.58} &\colorbox{green!25}{2.89} &4.06 &2.83\\
C2 &\colorbox{red!50}{3.07} &\colorbox{red!50}{1.98} &\colorbox{red!50}{2.90} &\colorbox{red!50}{2.04} &\colorbox{red!50}{2.82} &\colorbox{red!50}{1.70} &2.71 &2.01 &2.36 &1.44\\
C3 &2.39 &1.64 &2.44 &1.55 &2.74 &1.75 &2.41 &1.56 &2.71 &1.61\\
C4 &2.34 &1.91 &2.49 &1.82 &2.96 &1.81 &2.41 &1.63 &2.56 &1.71\\
C5 &2.39 &1.61 &2.93 &2.29 &\colorbox{red!50}{3.34} &\colorbox{red!50}{2.13} &2.89 &1.77 &2.58 &1.44\\
C6 &2.96 &2.10 &3.06 &2.04 &3.55 &2.16 &3.07 &2.26 &3.13 &2.13\\
C7 &3.25 &2.82 &3.77 &3.83 &3.89 &3.50 &3.62 &4.08 &3.78 &3.21\\
\hline
T1 &\colorbox{green!25}{4.31} &\colorbox{green!25}{4.40} &\colorbox{green!25}{4.42} &\colorbox{green!25}{4.25} &5.32 &7.59 &\colorbox{red!50}{6.36} &\colorbox{red!50}{6.20} &5.19 &4.64\\
T2 &11.31&10.72&\colorbox{green!25}{10.68}&\colorbox{green!25}{10.05}&16.22&17.43&15.26&12.28&14.19&14.76\\
   \hline
\end{tabular}
\vspace{0.75em}
\caption{Mean and standard deviation for Mean Angle Direction (MAD) metric in degrees for data corrected by Block-Matching, HySCO, TOPUP and Voss and uncorrected data.\colorbox{green!25}{light green} means that p-value shows significant  improvement with $ 10^{-3} < $ p-value $ < 5.10^{-2}$, \colorbox{red!50}{weak red} means that p-value shows significant deterioration. BS: brain stem.
\label{table:mean_mad}}
\end{center}
\end{table}

\begin{figure}
\begin{center}
  \includegraphics[width=\textwidth]{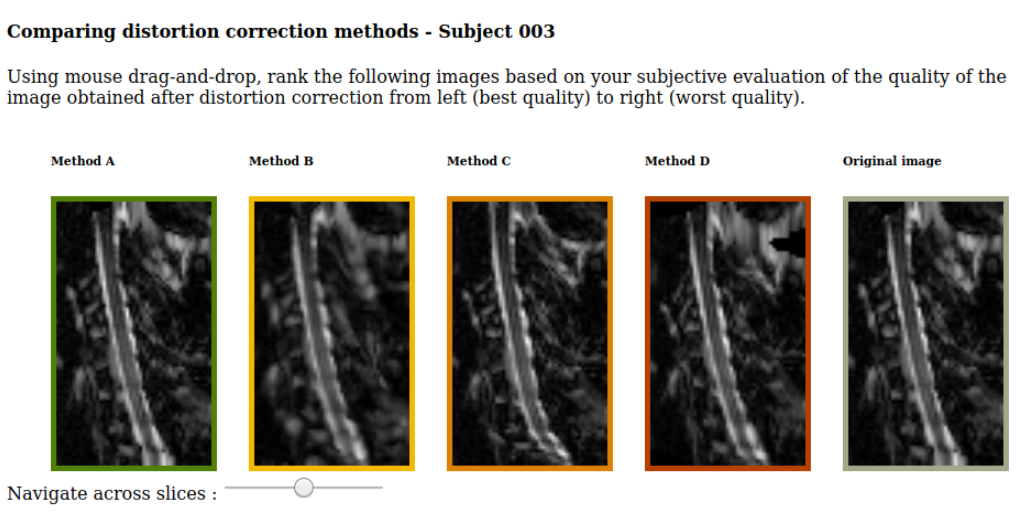}
  \caption{Screenshot of the web interface for expert ranking. The
images corrected using each of the four method are presented side-by-side,
along with the original image. The order in which the methods are displayed is
shuffled randomly for each subject, and the methods are labeled neutrally as
"Method~A", "Method~B", \ldots, etc. \label{fig:web-interface}}
\end{center}
\end{figure}

\begin{table}
\begin{center}
\begin{tabular}{|c|c|c|c|c|c|c|c|c|c|c|}
  \hline
\multicolumn{11}{c}{Angular Concentration of directions (ACD)}\\
  \hline
    ~&\multicolumn{2}{c|}{\textbf{Block-Matching}}&\multicolumn{2}{c|}{\textbf{HySCO}}&\multicolumn{2}{c|}{\textbf{TOPUP}}&\multicolumn{2}{c|}{\textbf{Voss}}&\multicolumn{2}{c|}{\textbf{Uncorrected}}\\
    \hline
    \textbf{Level}&Mean&STD&Mean&STD&Mean&STD&Mean&STD&Mean&STD\\
    \hline
    BS &\colorbox{green!100}{84.32} &\colorbox{green!100}{8.66} &\colorbox{green!25}{82.06} &\colorbox{green!25}{8.57} &78.10 &9.46 &\colorbox{green!100}{84.96} &\colorbox{green!100}{8.01} &80.22 &8.71\\
    \hline
C1 &\colorbox{green!25}{96.49} &\colorbox{green!25}{4.96} &\colorbox{green!25}{95.93} &\colorbox{green!25}{4.56} &95.60 &5.84 &\colorbox{green!25}{96.06} &\colorbox{green!25}{3.92} &95.10 &5.76\\
C2 &\colorbox{green!100}{97.75} &\colorbox{green!100}{2.14} &97.23 &2.29 &\colorbox{green!25}{97.68} &\colorbox{green!25}{2.20} &\colorbox{green!25}{97.57} &\colorbox{green!25}{2.21} &97.25 &2.55\\
C3 &97.91 &2.08 &97.80 &2.01 &97.82 &1.89 &97.83 &2.23 &97.71 &2.23\\
C4 &97.73 &2.34 &97.77 &1.82 &97.59 &2.33 &97.93 &1.57 &97.83 &1.69\\
C5 &97.62 &1.68 &97.35 &2.06 &\colorbox{green!25}{97.83} &\colorbox{green!25}{1.64} &\colorbox{green!25}{97.76} &\colorbox{green!25}{1.44} &97.48 &1.63\\
C6 &96.08 &4.03 &95.80 &3.75 &\colorbox{green!25}{96.53} &\colorbox{green!25}{3.56} &\colorbox{green!100}{96.57} &\colorbox{green!100}{3.19} &96.02 &3.31\\
C7 &94.63 &6.17 &\colorbox{green!25}{93.69} &\colorbox{green!25}{6.52} &94.84 &7.44 &\colorbox{green!25}{95.03} &\colorbox{green!25}{5.05} &94.32 &6.08\\
\hline
T1 &\colorbox{green!100}{93.09} &\colorbox{green!100}{7.92} &\colorbox{green!25}{91.37} &\colorbox{green!25}{7.49} &91.88 &9.17 &\colorbox{green!25}{91.61} &\colorbox{green!25}{7.70} &89.99 &9.58\\
T2 &\colorbox{green!100}{83.56} &\colorbox{green!100}{12.66}&80.23 &14.47&\colorbox{green!100}{85.00} &\colorbox{green!100}{13.99}&\colorbox{green!100}{84.67} &\colorbox{green!100}{11.40}&77.59 &16.15\\
  \hline
\end{tabular}
\vspace{0.75em}
\caption{Mean and standard deviation multiplied by $\mathbf {10^2}$ for ACD metric for data corrected by Block-Matching, HySCO, TOPUP and Voss and uncorrected data.\colorbox{green!100}{Dark green} means that p-value shows significant improvement with p-value $ < \mathrm 10^{-3}$,\colorbox{green!25}{light green} means that p-value shows also significant improvement but $ 10^{-3} < $ p-value $ < 5.10^{-2}$. BS: brain stem.
\label{table:mean_acd}}
\end{center}
\end{table}

\begin{table}[!htbp]
\begin{center}
\begin{tabular}{|c|c|c|c|}
  \hline
    & Voss & BM & TOPUP \\
     \hline
   HySCO & 41 vs 46 (p=0.72) & 49 vs 28 (p=0.0019)  & 76 vs 11 (p=$ 3.10^{-5}$)\\
   Voss & NA &59 vs28 (p=0.14) & 77 vs 10 (p=0.0027)\\
   BM & NA & NA & 67 vs 20 (p=0.01)\\
   \hline
\end{tabular}
\vspace{0.75em}
\caption{Summary of the pairwise logistic regression (accounting for a random rater effect): we report for each pair of methods the score for method 1 (row) vs method 2 (column). For instance, among the 87 subjective evaluations (29 subjects, 3 raters), Voss ranked behind HySCO 41 times. \label{table:subjective_table}}
\end{center}
\end{table}

\begin{table}[!htbp]
\begin{center}
\begin{tabular}{|c|c|c|c|c|c|}
  \hline
  \multicolumn{6}{c}{\textbf{Cross-correlation}}\\
  \hline
  \textbf{Methods}&BM&HySCO&TOPUP&Voss&Uncorrected\\
  \hline
  Mean &\colorbox{green!25}{0.185} &\colorbox{green!100}{0.242} & 0.179 & 0.148 & 0.160\\
  \hline
  STD&\colorbox{green!25}{0.114} &\colorbox{green!100}{0.119} & 0.150 & 0.115 & 0.107\\
  \hline
  P-value &\colorbox{green!25}{0.029}&\colorbox{green!100}{2.10$^{-4}$} & 0.189 & 0.173&-\\
  \hline
  t-statistic &\colorbox{green!25}{2.215} &\colorbox{green!100}{7.303} & 1.325 & -1.375 &-\\
  \hline
\end{tabular}
\vspace{0.75em}
\caption{Paired Tukey test for Block-Matching, HySCO, TOPUP and Voss for
Cross-correlation for the whole of the spinal cord region.
\colorbox{green!100}{Dark green} means that p-value shows significant
improvement and inferior to $ \mathrm 10^{-3}$, \colorbox{green!25}{light green}
means that p-value shows significant improvement but $ 10^{-3} < $ p-value $ <
5.10^{-2}$.
\label{table:cross_corr}}
\end{center}
\end{table}

\begin{table}[!htbp]
\begin{center}
\begin{tabular}{|c|c|c|c|c|c|}
  \hline
  \multicolumn{6}{c}{\textbf{Mutual information}}\\
  \hline
  \textbf{Methods}&BM&HySCO&TOPUP&Voss&Uncorrected\\
    \hline
  Mean & 0.304 & 0.330 & 0.312 & 0.293 & 0.277\\
  \hline
  STD& 0.051 & 0.049 & 0.064 & 0.053 & 0.054\\
  \hline
  P-value &\colorbox{green!100}{10$^{-5}$}&\colorbox{green!100}{4.10$^{-14}$} &\colorbox{green!100}{6.10$^{-7}$} &\colorbox{green!100}{9.10$^{-4}$} &-\\
  \hline
  t-statistic & 4.652 & 9.198 & 5.433 & 3.449 &-\\
  \hline
\end{tabular}
\vspace{0.75em}
\caption{Paired Tukey test for Block-Matching, HySCO, TOPUP and Voss for Mutual
information for the whole of the spinal cord region. \colorbox{green!100}{Dark
green} means that p-value shows significant improvement and inferior to $
\mathrm 10^{-3}$, \colorbox{green!25}{light green} means that p-value shows
significant improvement but $ 10^{-3} < $ p-value $ < 5.10^{-2}$.
\label{table:mutual_info}}
\end{center}
\end{table}

\begin{table}[!htbp]
\begin{center}
\begin{tabular}{|c|c|c|c|c|c|c|c|c|c|c|}
\hline
~&\multicolumn{2}{c|}{\textbf{BM}}&\multicolumn{2}{c|}{\textbf{HySCO}}&\multicolumn{2}{c|}{\textbf{TOPUP}}&\multicolumn{2}{c|}{\textbf{Voss}}&\multicolumn{2}{c|}{\textbf{Uncorrected}}\\
\hline
\textbf{\small Level}&Mean&STD&Mean&STD&Mean&STD&Mean&STD&Mean&STD\\
\hline
C1 & 560.5  & 123 & \colorbox{blue!20}{583.9} & 106 & 555.9  & 118 & \colorbox{blue!20}{555.8} & 110 & 568.7 & 104 \\
C2 & 530.7  & 118 & 552.0  & 101 & 528.7  & 123 & 526.8  & 103 & 538.7 & 108 \\
C3 & 913.2  & 155 & \colorbox{blue!20}{967.6} & 165 & 911.5  & 173 & 934.8  & 146 & 917.8 & 155 \\
C4 & 959.8  & 157 &\colorbox{blue!20}{ 1021}  & 163 & 923.0  & 175 & 949.4  & 142 & 943.4 & 152 \\
C5 & 855.4  & 158 & \colorbox{blue!20}{901.6} & 150 & 823.3  & 170 & 837.3  & 143 & 835.5 & 150 \\
C6 & \colorbox{blue!20}{777.3} & 137 & \colorbox{blue!20}{810.8} & 126 & 725.1  & 161 & 753.7  & 141 & 749.6 & 134 \\
C7 & 743.6  & 127 & \colorbox{blue!20}{764.4} & 120 & \colorbox{blue!20}{681.7} & 134 & \colorbox{blue!20}{710.4} & 132 & 732.6 & 127 \\
\hline
T1 & \colorbox{blue!20}{790.4} & 138 & \colorbox{blue!20}{820.9} & 153 & \colorbox{blue!20}{722.3} & 158 & 742.8  & 173 & 757.6 & 158 \\
T2 & \colorbox{blue!20}{456.9} & 277 & \colorbox{blue!20}{501.0} & 283 & 363.7  & 265 & \colorbox{blue!20}{339.5} & 283 & 397.9 & 271 \\
 \hline
\end{tabular}
\vspace{0.75em}
\caption{Volume of every vertebral level (in mm$^3$) of spinal cord, segmented
mask, corrected by Block-Matching, HySCO, TOPUP, Voss and uncorrected. Significant differences ($p<0.05$) are indicated by blue color.
\label{table:volume_level}}
\end{center}
\end{table}

\newpage
\bibliography{biblio}

\begin{thebibliography}{32}
\providecommand{\natexlab}[1]{#1}
\providecommand{\url}[1]{\texttt{#1}}
\providecommand{\urlprefix}{}

\bibitem[{Stehling et~al.(1991)Stehling, Michael K and Turner, Robert and
  Mansfield, Peter}]{stehling1991echo}
Stehling MK, Turner R, Mansfield P.
\newblock Echo-planar imaging: magnetic resonance imaging in a fraction of a
  second.
\newblock Science 1991;254(5028):43--50.

\bibitem[{Jezzard and Balaban(1995)Jezzard, Peter and Balaban, Robert
  S}]{jezzard1995correction}
Jezzard P, Balaban RS.
\newblock Correction for geometric distortion in echo planar images from B0
  field variations.
\newblock Magnetic resonance in medicine 1995;34(1):65--73.

\bibitem[{Macdonald and Ruthotto(2016)Macdonald, Jan and Ruthotto,
  Lars}]{macdonald2016efficient}
Macdonald J, Ruthotto L.
\newblock Efficient numerical optimization for susceptibility artifact
  correction of EPI-MRI.
\newblock arXiv preprint arXiv:160700531 2016;.

\bibitem[{Saritas et~al.(2014)Saritas, Emine U and Holdsworth, Samantha J and
  Bammer, Roland}]{saritas2014susceptibility}
Saritas EU, Holdsworth SJ, Bammer R.
\newblock Susceptibility artifacts.
\newblock In: Cohen-Adad J, Wheeler-Kingshott C, editors. Quantitative MRI of
  the Spinal Cord Elsevier; 2014.p. 91--105.

\bibitem[{Chang and Fitzpatrick(1992)Chang, Hsuan and Fitzpatrick, J
  Michael}]{chang1992technique}
Chang H, Fitzpatrick JM.
\newblock A technique for accurate magnetic resonance imaging in the presence
  of field inhomogeneities.
\newblock IEEE Transactions on medical imaging 1992;11(3):319--329.

\bibitem[{Reber et~al.(1998)Reber, Paul J and Wong, Eric C and Buxton, Richard
  B and Frank, Lawrence R}]{reber1998correction}
Reber PJ, Wong EC, Buxton RB, Frank LR.
\newblock Correction of off resonance-related distortion in echo-planar imaging
  using EPI-based field maps.
\newblock Magnetic Resonance in Medicine 1998;39(2):328--330.

\bibitem[{Andersson et~al.(2003)Andersson, Jesper LR and Skare, Stefan and
  Ashburner, John}]{andersson2003correct}
Andersson JL, Skare S, Ashburner J.
\newblock How to correct susceptibility distortions in spin-echo echo-planar
  images: application to diffusion tensor imaging.
\newblock Neuroimage 2003;20(2):870--888.

\bibitem[{Holland et~al.(2010)Holland, Dominic and Kuperman, Joshua M and Dale,
  Anders M}]{holland2010efficient}
Holland D, Kuperman JM, Dale AM.
\newblock Efficient correction of inhomogeneous static magnetic field-induced
  distortion in Echo Planar Imaging.
\newblock Neuroimage 2010;50(1):175--183.

\bibitem[{Robson et~al.(1997)Robson, Matthew D and Gore, John C and Constable,
  R Todd}]{robson1997measurement}
Robson MD, Gore JC, Constable RT.
\newblock Measurement of the point spread function in MRI using constant time
  imaging.
\newblock Magnetic resonance in medicine 1997;38(5):733--740.

\bibitem[{Zeng and Constable(2002)Zeng, Huairen and Constable, R
  Todd}]{zeng2002image}
Zeng H, Constable RT.
\newblock Image distortion correction in EPI: comparison of field mapping with
  point spread function mapping.
\newblock Magnetic Resonance in Medicine: An Official Journal of the
  International Society for Magnetic Resonance in Medicine 2002;48(1):137--146.

\bibitem[{Zaitsev et~al.(2004)Zaitsev, M and Hennig, J and Speck,
  O}]{zaitsev2004point}
Zaitsev M, Hennig J, Speck O.
\newblock Point spread function mapping with parallel imaging techniques and
  high acceleration factors: fast, robust, and flexible method for echo-planar
  imaging distortion correction.
\newblock Magnetic Resonance in Medicine: An Official Journal of the
  International Society for Magnetic Resonance in Medicine
  2004;52(5):1156--1166.

\bibitem[{Speck et~al.(2008)Speck, Oliver and Stadler, J and Zaitsev,
  M}]{speck2008high}
Speck O, Stadler J, Zaitsev M.
\newblock High resolution single-shot EPI at 7T.
\newblock Magnetic Resonance Materials in Physics, Biology and Medicine
  2008;21(1-2):73.

\bibitem[{Lundell et~al.(2013)Lundell, Henrik and Barthelemy, Dorothy and
  Biering-S{\o}rensen, Fin and Cohen-Adad, Julien and Nielsen, Jens Bo and
  Dyrby, Tim B}]{lundell2013fast}
Lundell H, Barthelemy D, Biering-S{\o}rensen F, Cohen-Adad J, Nielsen JB, Dyrby
  TB.
\newblock Fast diffusion tensor imaging and tractography of the whole cervical
  spinal cord using point spread function corrected echo planar imaging.
\newblock Magnetic resonance in medicine 2013;69(1):144--149.

\bibitem[{Cohen-Adad et~al.(2009)Cohen-Adad, J and Lundell, Henrik and
  Rossignol, S}]{cohen2009distortion}
Cohen-Adad J, Lundell H, Rossignol S.
\newblock Distortion correction in spinal cord DTI: what's the best approach.
\newblock In: Proceedings of the 17th Annual Meeting of ISMRM, Honolulu, USA,
  vol. 3178; 2009. .

\bibitem[{Hedouin et~al.(2017)Hedouin, Renaud and Commowick, Olivier and
  Bannier, Elise and Scherrer, Benoit and Taquet, Maxime and Warfield, Simon K
  and Barillot, Christian}]{hedouin2017block}
Hedouin R, Commowick O, Bannier E, Scherrer B, Taquet M, Warfield SK, et~al.
\newblock Block-Matching Distortion Correction of Echo-Planar Images With
  Opposite Phase Encoding Directions.
\newblock IEEE Trans Med Imaging 2017;36(5):1106--1115.

\bibitem[{Ruthotto et~al.(2012)Ruthotto, L and Kugel, H and Olesch, J and
  Fischer, B and Modersitzki, J and Burger, M and Wolters,
  CH}]{ruthotto2012diffeomorphic}
Ruthotto L, Kugel H, Olesch J, Fischer B, Modersitzki J, Burger M, et~al.
\newblock Diffeomorphic susceptibility artifact correction of
  diffusion-weighted magnetic resonance images.
\newblock Physics in Medicine \& Biology 2012;57(18):5715.

\bibitem[{Voss et~al.(2006)Voss, Henning U and Watts, Richard and Ulu{\u{g}},
  Aziz M and Ballon, Doug}]{voss2006fiber}
Voss HU, Watts R, Ulu{\u{g}} AM, Ballon D.
\newblock Fiber tracking in the cervical spine and inferior brain regions with
  reversed gradient diffusion tensor imaging.
\newblock Magnetic resonance imaging 2006;24(3):231--239.

\bibitem[{Ourselin et~al.(2000)Ourselin, S{\'e}bastien and Roche, Alexis and
  Prima, Sylvain and Ayache, Nicholas}]{ourselin2000block}
Ourselin S, Roche A, Prima S, Ayache N.
\newblock Block matching: A general framework to improve robustness of rigid
  registration of medical images.
\newblock In: International Conference on Medical Image Computing And
  Computer-Assisted Intervention Springer; 2000. p. 557--566.

\bibitem[{Commowick et~al.(2012)Commowick, Olivier and Wiest-Daessl{\'e},
  Nicolas and Prima, Sylvain}]{commowick2012automated}
Commowick O, Wiest-Daessl{\'e} N, Prima S.
\newblock Automated diffeomorphic registration of anatomical structures with
  rigid parts: Application to dynamic cervical MRI.
\newblock In: International Conference on Medical Image Computing and
  Computer-Assisted Intervention Springer; 2012. p. 163--170.

\bibitem[{Avants et~al.(2008)Avants, Brian B and Epstein, Charles L and
  Grossman, Murray and Gee, James C}]{avants2008symmetric}
Avants BB, Epstein CL, Grossman M, Gee JC.
\newblock Symmetric diffeomorphic image registration with cross-correlation:
  evaluating automated labeling of elderly and neurodegenerative brain.
\newblock Medical image analysis 2008;12(1):26--41.

\bibitem[{Cohen-Adad et~al.(2008)Cohen-Adad, J and Descoteaux, M and Rossignol,
  S and Hoge, R D and Deriche, R and Benali, H}]{cohen2008detection}
Cohen-Adad J, Descoteaux M, Rossignol S, Hoge RD, Deriche R, Benali H.
\newblock Detection of multiple pathways in the spinal cord using q-ball
  imaging.
\newblock Neuroimage 2008;42(2):739--749.

\bibitem[{De~Leener et~al.(2014)De Leener, Benjamin and Kadoury, Samuel and
  Cohen-Adad, Julien}]{de2014robust}
De~Leener B, Kadoury S, Cohen-Adad J.
\newblock Robust, accurate and fast automatic segmentation of the spinal cord.
\newblock Neuroimage 2014;98:528--536.

\bibitem[{De~Leener et~al.(2015)De Leener, Benjamin and Cohen-Adad, Julien and
  Kadoury, Samuel}]{de2015automatic}
De~Leener B, Cohen-Adad J, Kadoury S.
\newblock Automatic segmentation of the spinal cord and spinal canal coupled
  with vertebral labeling.
\newblock IEEE transactions on medical imaging 2015;34(8):1705--1718.

\bibitem[{Mardia and Jupp(2009)Mardia, Kanti V and Jupp, Peter
  E}]{mardia2009directional}
Mardia KV, Jupp PE.
\newblock Directional statistics, vol. 494.
\newblock John Wiley \& Sons; 2009.

\bibitem[{Mohammadi et~al.(2013)Mohammadi, Siawoosh and Freund, Patrick and
  Feiweier, Thorsten and Curt, Armin and Weiskopf,
  Nikolaus}]{mohammadi2013impact}
Mohammadi S, Freund P, Feiweier T, Curt A, Weiskopf N.
\newblock The impact of post-processing on spinal cord diffusion tensor
  imaging.
\newblock Neuroimage 2013;70:377--385.

\bibitem[{Xu et~al.(2013)Xu, J and Shimony, J S and Klawiter, E C and Snyder, A
  Z and Trinkaus, K and Naismith, R T and Benzinger, T LS and Cross, A H and
  Song, SK}]{xu2013improved}
Xu J, Shimony JS, Klawiter EC, Snyder AZ, Trinkaus K, Naismith RT, et~al.
\newblock Improved in vivo diffusion tensor imaging of human cervical spinal
  cord.
\newblock Neuroimage 2013;67:64--76.

\bibitem[{De~Leener et~al.(2017)De Leener, B and L{\'e}vy, S and Dupont, S M
  and Fonov, V S and Stikov, N and Collins, D L and Callot, V and Cohen-Adad,
  J}]{de2017sct}
De~Leener B, L{\'e}vy S, Dupont SM, Fonov VS, Stikov N, Collins DL, et~al.
\newblock SCT: Spinal Cord Toolbox, an open-source software for processing
  spinal cord MRI data.
\newblock Neuroimage 2017;145:24--43.

\bibitem[{De~Leener et~al.(2018)De Leener, Benjamin and Fonov, Vladimir S and
  Collins, D Louis and Callot, Virginie and Stikov, Nikola and Cohen-Adad,
  Julien}]{de2018pam50}
De~Leener B, Fonov VS, Collins DL, Callot V, Stikov N, Cohen-Adad J.
\newblock PAM50: Unbiased multimodal template of the brainstem and spinal cord
  aligned with the ICBM152 space.
\newblock NeuroImage 2018;165:170--179.

\bibitem[{L{\'e}vy et~al.(2015)L{\'e}vy, Simon and Benhamou, M and Naaman, C
  and Rainville, Pierre and Callot, Virginie and Cohen-Adad,
  Julien}]{levy2015white}
L{\'e}vy S, Benhamou M, Naaman C, Rainville P, Callot V, Cohen-Adad J.
\newblock White matter atlas of the human spinal cord with estimation of
  partial volume effect.
\newblock Neuroimage 2015;119:262--271.

\bibitem[{Snoussi et~al.(2017)Snoussi, H and Caruyer, E and Commowick, O and
  Bannier, E and Barillot, C}]{snoussi2017comparison}
Snoussi H, Caruyer E, Commowick O, Bannier E, Barillot C.
\newblock Comparison of inhomogeneity distortion correction methods in
  diffusion MRI of the spinal cord.
\newblock In: ESMRMB-34th Annual Scientific Meeting; 2017. .

\bibitem[{Dauleac et~al.(2021)Dauleac, Corentin and Bannier, Elise and Cotton,
  Fran{\c{c}}ois and Frindel, Carole}]{dauleac2021effect}
Dauleac C, Bannier E, Cotton F, Frindel C.
\newblock Effect of distortion corrections on the tractography quality in
  spinal cord diffusion-weighted imaging.
\newblock Magnetic Resonance in Medicine 2021;.

\bibitem[{Hu et~al.(2020)Hu, Zhangxuan and Wang, Yishi and Zhang, Zhe and
  Zhang, Jieying and Zhang, Huimao and Guo, Chunjie and Sun, Yuejiao and Guo,
  Hua}]{hu2020distortion}
Hu Z, Wang Y, Zhang Z, Zhang J, Zhang H, Guo C, et~al.
\newblock Distortion correction of single-shot EPI enabled by deep-learning.
\newblock NeuroImage 2020;221:117170.

\end{thebibliography}
\end{document}